\documentclass[final,authoryear,11pt]{elsarticle} 

\usepackage{graphicx,amssymb,amsfonts,amsthm}
\usepackage[tbtags]{amsmath}
\usepackage{natbib}
\usepackage[utf8]{inputenc}
\usepackage[british]{babel}
\usepackage{hyperref}
\usepackage[mathscr]{eucal}
\usepackage[toc,page]{appendix}
\usepackage{setspace}

\pdfoutput=1

\biboptions{numbers}
\journal{Journal of Marine Systems}

\newcommand{\pbi}{\begin{itemize}}
\newcommand{\pei}{\end{itemize}}

\begin{document}

\begin{frontmatter}

\title{Evaluating Environmental Joint Extremes for the Offshore Industry}

\author{Kevin Ewans}
\cortext[cor1]{kevin.ewans@shell.com, Tel:+6085453498}
\address{Sarawak Shell Bhd., 50450 Kuala Lumpur, Malaysia.}
\author{Philip Jonathan}
\address{Shell Projects and Technology Thornton, CH1 3SH, UK.}

\begin{abstract}
Understanding extreme ocean environments and their interaction with fixed and floating structures is critical for the design of offshore and coastal facilities. The joint effect of various ocean variables on extreme responses of offshore structures is fundamental in determining the design loads. For example, it is known that mean values of wave periods tend to increase with increasing storm intensity, and a floating system responds in a complex way to both variables.
However, specification of joint extremes in design criteria has often been somewhat \textit{ad hoc}, being based on fairly arbitrary combinations of extremes of variables estimated independently. Such approaches are even outlined in design guidelines. Mathematically more consistent estimates of the joint occurrence of extreme environmental variables fall into two camps in the offshore industry -- response-based and response-independent. Both are outlined here, with emphasis on response-independent methods, particularly those based on the conditional extremes model recently introduced by \cite{HffTwn04}, which has a solid theoretical motivation. Several applications using the new methods are presented.
\\

\textbf{Keywords}: offshore design; floating structures; joint extremes; conditional extremes; covariates;\\
\end{abstract}

\end{frontmatter}


\section{Introduction}\label{Sct1}
Offshore structures must be designed to very low probabilities of failure due to storm loading. Design codes stipulate that offshore structures should be designed to exceed specific levels of reliability, expressed in terms of an annual probability of failure or return-period. This requires specification of values of environmental variables with very low probabilities of occurrence. More specifically, since the goal is to determine structural loading due to environmental forcing, it is the combination of environmental phenomena with a given return-period that is sought. For example, most physical systems respond to environmental conditions in a manner that cannot be represented by a single variable - the pitch of a vessel is as much a function of the wave period or wave length as it is of the wave height, and it is necessary to also specify appropriate associated values of period for a given extreme wave height.

The goal is thus to design an offshore facility to withstand extreme environmental conditions that will occur during its lifetime with an appropriate optimum risk level. The level of risk is set by weighing the consequences of failure against the cost of over-designing. Facilities with a 20 to 30 year lifetime generally use 100-year metocean criteria, which with typical implicit and explicit safety factors, leads to annual probabilities of failure of $10^{-3}$ to $10^{-5}$. For example, the load-resistance factor design (ISO, API) have an environmental load factor, $\gamma_E=$1.35, to use with the loading calculated from an appropriate combination of environmental variables with a return-period of 100 years. The challenge is the choice of the appropriate combination of environmental variables through extreme value analyses.

Estimation of the extremes of single variables is relatively straight forward, given a long time series or time history of that variable that spans many years, and as a consequence, combinations of independently derived variables are often used for estimating environmental forces. One could for example, use the maximum wave height, wind speed, and current speed each with a return period of 100 years to derive the environmental loading with a return period of 100 years, but unless the winds, waves, and currents are perfectly correlated, the probability of this combination of variables is considerably less than 0.01 per annum. Some design codes and guidelines, would suggest taking the 100 year return period of one variable together with the value of an associated variable for a shorter return period. For example, the DNV recommended practice for on-bottom stability of pipelines suggests the combination of the 100-year return condition for waves combined with the 10-year return condition for current or vice-versa, when detailed information about the joint probability of waves and current is not available. Without prior knowledge, the direction of the winds, waves, and currents can even be considered to be the same.

The simple combination of independent variables also glosses over the diverse climates that characterise the World's oceans. For example, the extreme meteorological phenomena in the Gulf of Mexico and the northwest coast of Australia are hurricanes. These are characterised by waves and currents that are driven by the local wind field, and there is a high probability of experiencing extreme winds, waves and currents together. In the Gulf of Guinea, extreme wave events are associated with swells from South Atlantic storms. The swells run normal to coast, while the currents from ocean circulation run along coast, and are independent from the swell. Accordingly, the probability of experiencing extreme waves and extreme currents is low, and the probability that the waves and currents are collinear is even smaller. In the Arabian Gulf, the wave extremes are due to the Shamal, whereas the currents are dominated by tides. As a result, and like the Gulf of Guinea case, the probability of experiencing extreme waves and extreme currents together is relatively low, but unlike the Gulf of Guinea, they are largely inline.

It is therefore clear that simple and relatively arbitrary combinations of independent criteria will result in joint criteria with an unknown probability, and further a given choice of combination will result in joint criteria with different probabilities for different oceans, when in fact the desired outcome are conditions that will give facilities designed to the same level of reliability. Accordingly, joint criteria with known probability of occurrence are required.

The approaches used by the offshore industry to calculate joint extreme environmental conditions essentially fall into two camps -- response-based and response-independent. The response-based approach relies on the specification of a response model giving load as a function of environment and permitting a back calculation of the environmental variables once an extreme load has been established. The response-independent or environmental approach involves developing joint criteria for the environmental variables alone associated with rare return periods.
A outline of the response-based approach is given in Section \ref{Sct2}. The main section of this paper is therefore Section \ref{Sct3}, which gives a review of contemporary methods for calculating joint extreme environmental variables. Discussion and summary are presented in Section \ref{Sct4}.
%

\section{Response--based methods} \label{Sct2}
Response-based methods involve calculating a key response or several key responses via a response function in which the variables are environmental variables. For example, \cite{TrmVnd95} describe generic response functions for the mud-line base shear and over-turning moment of steel jacket structures. Their response function is given in terms of a sum of terms involving variables of the winds, waves, and currents. The coefficients of the terms are determined by calibration of large number of conditions with a given wave kinematics and current profile on a one meter diameter vertical column.

With a given response function, a long-term data set of environmental variables can be converted into an equivalently long-term data set of responses, allowing an extreme value analysis of the response variable to be undertaken. Estimates of the extremes of the response variable can be made to a given annual probability of exceedence or return-period, and this value can be used in the original response function to back-calculate the environmental variables, to establish an appropriate design set of environmental variables for detailed engineering design. It should be noted that the response variable calculated from the response function need not be an actual engineering response or load, but it must have the same statistical behaviour with the environmental variables as an actual engineering response or load.

The back-calculation of the environmental variables from the response function is not trivial. In its simplest form the back-calculation involves establishing an optimum combination of environmental variables, based on relationships established from the data and assumptions that these relationships will also apply in the extreme. Usually, one of the variables, such as the wave height in the case of the steel jacket response functions, is assumed to be dominant and the value of this variable is set at the return-period of interest. The other variables are then determined from their respective relationships for this value of the dominant variable. The optimum set of variables when substituted in the response function would give the extreme value of the response variable. The optimum choice of variables can also be determined by extending a response-independent method with the addition of the response variable. Distributions of the values of the environmental variables conditional on an extreme response variable can then be established, and an appropriate choice such as the most probable of each variable can be made.
A consequence of the response-based approach, and in particular to having the probability distribution for the response or load variable, is that it is possible in principle to calculate the reliability of a structure against failure due to the environmental loading. We assume that the structural strength or resistance, $R$, of a structure can be characterised by a probability density function $f_R$. For any given value $x$ of structural resistance, the structure fails if the environmental load $E$ exceeds $x$. Writing the probability $\Pr(E>x)$ as $\bar{F}_E(x)$, it follows that the probability $p_F$ of structural failure is
\[ p_F = \int \bar{F}_E(x) f_R(x) dx \]
The estimation of reliability is central to the FORM and SORM methods (e.g. \citealt{WntEng98}). For a set of environmental variables $X$, a safety margin function $g(X)$ is defined such that when $g(X) \leq 0$, the structure will fail, otherwise it is safe. In the case of the steel jacket structure discussed above, $g(X)=R-E$; i.e., the structure will fail when the environmental load is greater than the resistance. The probability of failure is then determined from
\[ p_F=\int_{g(x) \leq 0} f(x) dx \]
where $f(x)$ is now the joint probability density function of the set of environmental variables. The integral is difficult to solve since both $f(x)$ and the integration boundary, $g(x)=0$, the failure surface, are multidimensional and usually nonlinear. The problem is simplified by expressing the $X$ as the product of (independent) conditional random variables $\tilde{X}_1, \tilde{X}_2, ...$. For example, for appropriately ordered variables we can write
\[\tilde{X}_1=X_1 \text{, then } \tilde{X}_2=X_2 | X_1 \text{ and } \tilde{X}_3=X_3 | X_1, X_2 \text{ and so on} \]
with cumulative distribution functions $F_{\tilde{X}_1}, F_{\tilde{X}_2}, ...$. These independent random variables can now be transformed in turn to standard normal random variables $U_1, U_2, ...$ via the probability integral or Rosenblatt transform
\[ F_{\tilde{X}_j}(x) = \Phi(u_j) \text{ for } j=1,2,...\]
where $\Phi$ is the cumulative distribution function of the standard normal distribution. The probability of failure is then evaluated using
\[ p_F=\int_{g_U(u) \leq 0} \phi(u) du \]
for transformed failure surface $g_U$, where $\phi$ is the probability density function of a set of independent standard normal random variables. Thus, the contours of constant probability density of the integrand are concentric circles (bivariate case) or hyper-spheres in higher dimensions.
To facilitate solution, the integration boundary $g_U(u)=0$ is simplified by truncating its Taylor expansion about an as yet unknown point, $u^*$, to first order (FORM) or second order (SORM). $u^*$ is the point that has the highest probability density on $g_U(u)=0$, to minimise accuracy loss (the integrand function quickly diminishes away from the expansion point), and is referred to as the Most Probable Point (MPP).

The MPP is found by minimising $\|u\|$ for $g_U(u)=0$,  the minimum distance from origin to the failure surface. The minimum distance $\beta=\|u^*\|$ is called the reliability index, and the probability of failure is now simply $p_F=1-\phi(\beta)$. The value of $u^*$ can be transformed back to a corresponding $x^*$ (in terms of the original variables) to establish the failure design set.
%

\section{Response-independent methods} \label{Sct3}
Response-independent methods for establishing combinations of environmental variables for design require joint distributions that describe the behaviour of the variables when one or more is extreme to be established directly from the environmental variables themselves. A particular combination of variables with a given low probability of occurrence can then be specified. Reference to a response variable is not required but could be used to further optimise selection of variables. In this sense, response-based methods are only different in that they involve finding the most likely combination of environmental variables to produce a target response value.

In the case of FORM or SORM, the failure surface is the target and a failure probability is calculated, but conversely if the target is a failure probability, a design point can be calculated on an associated failure surface. \cite{WntEA93} demonstrate this approach, which they refer to as inverse FORM, to calculate probability contours of joint occurrences of environmental variables. The design point, $u^*$, is found by minimising $g_U(u)$ for $\|u\|=\beta$. The FORM and SORM failure surfaces are tangential to the contour at $u^*$, but for design the behaviour of the system can be checked to ensure the actual failure surface is outside the contour for that probability.

An example of the application of inverse FORM is given in Figure 1. The plot shows contours of the joint probability of the significant wave height, $H_S$, and spectral period, $T_P$, following the joint probability model proposed by \cite{HvrNhs86}.
\begin{figure}[p]
  \center\includegraphics[width=1\textwidth]{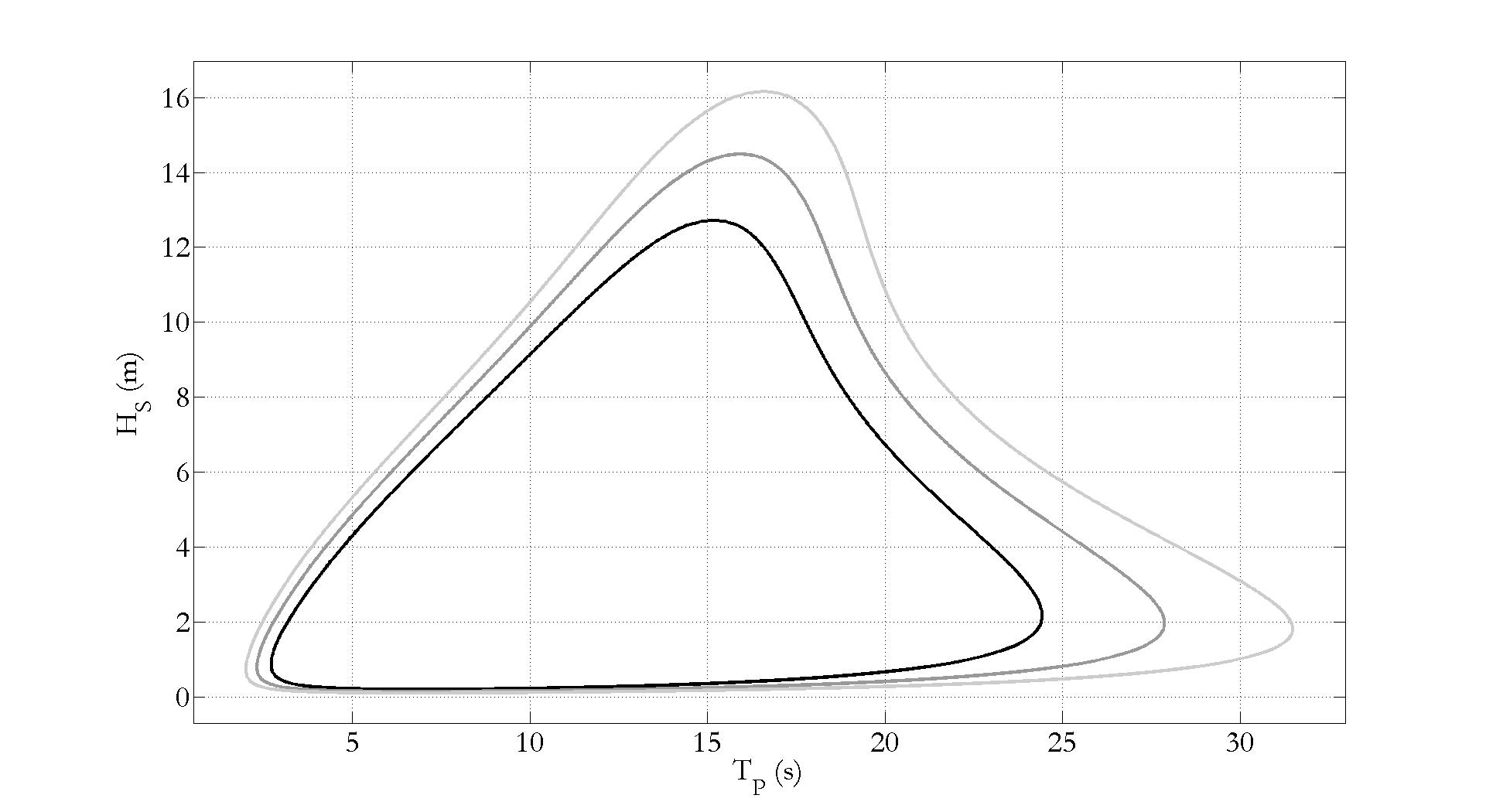}\\
  \caption{Contours of the joint occurrence of significant wave height and spectral peak period, for annual exceedence probabilities corresponding to 10-year (black), 100-year (dark grey), and 1000-year (light grey) return periods.}
  \label{Fig1}
\end{figure}

The inverse FORM approach requires us again to express the set of environmental variables in terms of a product of independent random variables. In the bivariate case, we might model the distribution of $X_1$ and $X_2 | X_1$, if there is good physical justification for doing so. In the model of \cite{HvrNhs86}, $H_S$ is modelled with a Weibull distribution and $T_P|H_S$ is modelled with a log-normal distribution. These model forms are motivated by good fitting performance to the body of a sample of data, but their validity for extremes is not known. In addition, inverse FORM is difficult to model beyond two variables, requiring a model for the probability density of a variable conditional on the occurrence of the others. In the case of three variables, the objective is to estimate $f(X_1,X_2,X_3 )$ with
\[ f(X_3,X_2,X_1 )=f(X_3|X_2,X_1)f(X_2|X_1)f(X_1) \]
but the difficulty lies in justifying the choice $X_3|X_2,X_1$ on physical grounds, and then estimating $f(X_3|X_2,X_1)$ - which is not straightforward.

The motivation for application of asymptotic distributions in extreme value analysis is the remarkable concept of max-stability. It can be shown that (appropriately shifted and scaled versions of) maxima from a large class of (``max-stable'') probability distributions have very similar statistical characteristics and the same distributional form.  As a result, in the univariate case, there is justification for modelling extremes of block maximum data (e.g. monthly maxima) with a generalised extreme value distribution and peaks over threshold data with a generalised Pareto distribution. However, in the multi-dimensional case, max-stability is only possible when (often unrealistic) component-wise maxima assumptions are appropriate. Nevertheless, the max-stable concept has been used for spatial extremes, with implicit asymptotic dependence assumed (\citealt{JntEwn13}).

The conditional extremes model of \cite{HffTwn04} provides a more general framework based on a (more realistic) limit assumption. It involves modelling the conditional distribution of one variable when the value of the conditioning variate is large, but a distinct advantage over a typical FORM analysis is that no prior knowledge of the forms of the distributions is required. Instead, asymptotic distributional forms are used.

The method is most clearly and most easily described in the case of two variables $(X,Y)$ but can be trivially extended to multi-dimensions. The marginal distribution of each variable is expressed on a Gumbel scale, by modelling variables in turn using a generalised Pareto distribution (assuming threshold exceedences) and then transforming using the probability integral transform. A parametric form then applies for the conditional distribution of one variable given large value of other
\[ (Y|X=x)=\alpha x+x^\beta Z \text{ for } x>u \]
for an appropriate threshold $u$, where $a\in(0,1]$ is the scale parameter, $b\in(-\infty,1]$ is the shape parameter, and $Z$ is a random variable, independent of $X$, converging with increasing $x$ to a non-degenerate limiting distribution, $G$ (which is assumed Gaussian for model fitting purposes only). In application, $Z$ is estimated from the residuals
\[ \hat{z}_i=\frac{y_i-\hat{\alpha} x_i}{x_i^{\hat{\beta}}} \text{ for } i=1,2,...\]
Then estimates of conditional extremes of $Y$ given $X$ are obtained by simulation by

\pbi
\item Drawing a threshold exceedence value $x$ of $X$ randomly from its standard Gumbel distribution,
\item Drawing a value $z$ of $Z$ randomly from the set of estimated values of $\hat{z}$,
\item Calculating $(y|x)=\hat{\alpha} x_i+ x_i^{\hat{\beta}} Z$, and finally
\item Transforming the pair $(x,y)$ from Gumbel to original physical scale using the probability integral transform.
\pei

\begin{figure}[p]
  \center\includegraphics[width=1\textwidth]{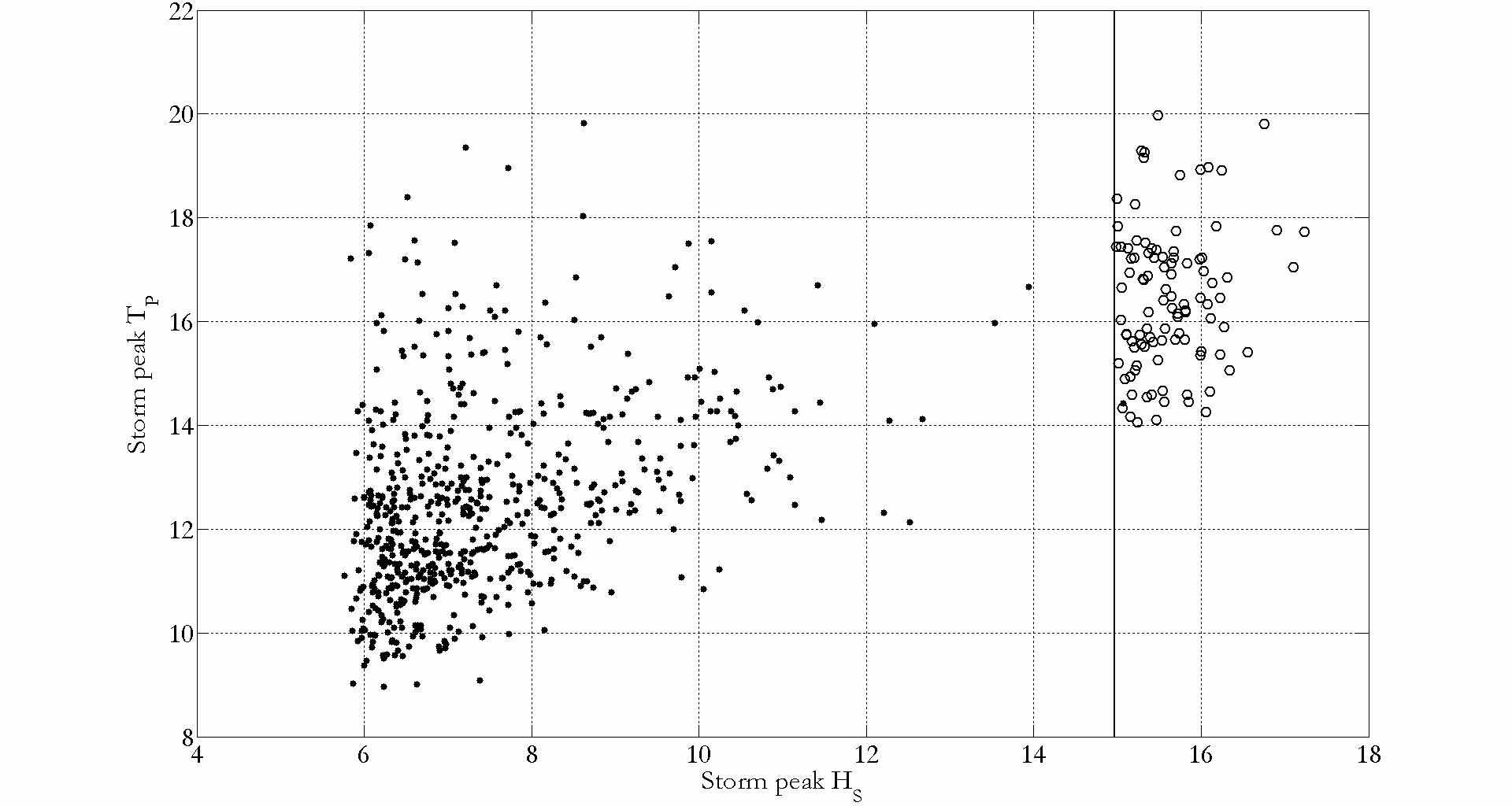}\\
  \caption{Measured storm peak significant wave height and associated spectral peak period from the northern North Sea (filled dots), and simulations of $H_S$ and conditional $T_P$ values for storm peak $H_S >15$m following conditional extremes modelling.}
  \label{Fig2}
\end{figure}

By way of example, an application of the Heffernan and Tawn method to wave data from several locations is given in \cite{JntFlnEwn10}. Figure \ref{Fig2} is a plot of measured storm peak significant wave height and associated spectral peak period from measurements in the northern North Sea, together with estimates from simulations for $H_S > 15$m from conditional extremes modelling. The plot shows a generally increasing trend in $T_P$ with $H_S$. The most probable value of $T_P$, which appears to be between 16s and 17s for $H_S > 15$m, is significantly less than the longest in the measured data.

An example of an application of the Heffernan and Tawn method to multi-dimensional problems is given by \cite{JntEwnFln12}, in which current profiles measured on the northwest shelf of Australia were analysed to derive extreme profiles with depth. Two and half years of current measurements, including both speed and direction, were made at eight depths through the water were available for the analysis. The steps in the analysis involve
\pbi
\item Resolving currents into major and minor axes of total current at each depth,
\item For each axes, separating tidal and residual components by a local harmonic analysis,
\item Calculating hourly maxima for each of the tidal and residual components, with the residual maxima to be used for the extreme value analysis, and the tidal maxima to be used for recombining with the residual simulations from conditional extremes modelling,
\item Applying the conditional extremes model to the residual hourly extremes
\pbi
\item fitting marginals with a generalised Pareto distribution,
\item transforming to Gumbel marginal scale,
\item fitting a multi-dimensional conditional extremes model (for all residual components) of the form
\[ (\mathbf{Y}_{[-k]} | Y_k=y_k)=\mathbf{\alpha}_k y_k+y_k^{\mathbf{\beta}_k} \mathbf{Z}_k \]
where $[-k]$ implies ``all except $k$'', bold characters indicate vectors, and componentwise multiplication assumed,
\pei
\item Simulating samples of joint extremes, where
\pbi
\item tidal components are re-sampled with replacement, and
\item sampled tidal components and residuals are added to provide hourly estimates of hourly maxima and minima along the major and minor axes.
\pei
\pei

\begin{figure}[p]
  \center\includegraphics[width=1\textwidth]{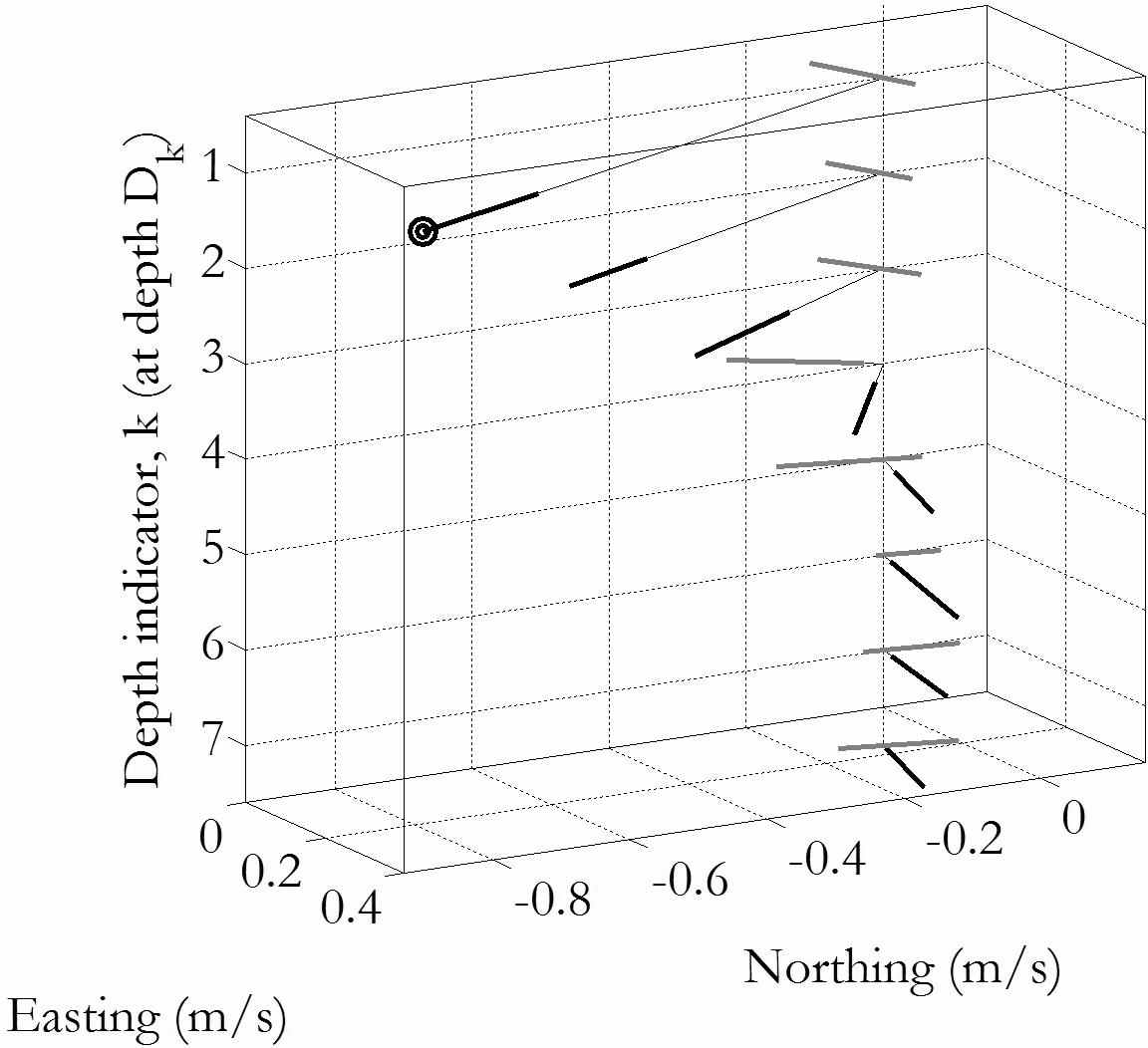}\\
  \caption{Maxima hourly extremes conditioned on exceedances of the 10-year return period current values at level D1.}
  \label{Fig3}
\end{figure}

An example of the results is given in Figure \ref{Fig3}, which shows median maxima hourly extremes conditioned on exceedances of the 10-year return period current values at depth D1. The figure suggests that the minor axis conditional extremes are approximately symmetric about zero at depths D1 to D3. At depth D4 however, there is systematic rotation of current components in a clockwise direction, with respect to axis directions defined using unconditioned sample at this depth. At depths D5 to D8, this trend is reversed; rotation is in an anti-clockwise direction.

The importance of accounting for covariates in univariate extreme value analyses has been demonstrated by \cite{JntEwnFrr08a}. The conditional extremes model can also be extended to include covariates in a relatively straight forward manner. The objective becomes to model the distribution of $T_P$ (say) when $H_S$ is extreme, as a function of storm direction $\theta$ as covariate, for which the conditional extremes model form becomes
\[ (T_P |H_S=h,\Theta=\theta)=\alpha_\theta h+h^{\beta_\theta} (\mu_\theta+\sigma_\theta Z) \]

\begin{figure}[p]
  \center\includegraphics[width=1\textwidth]{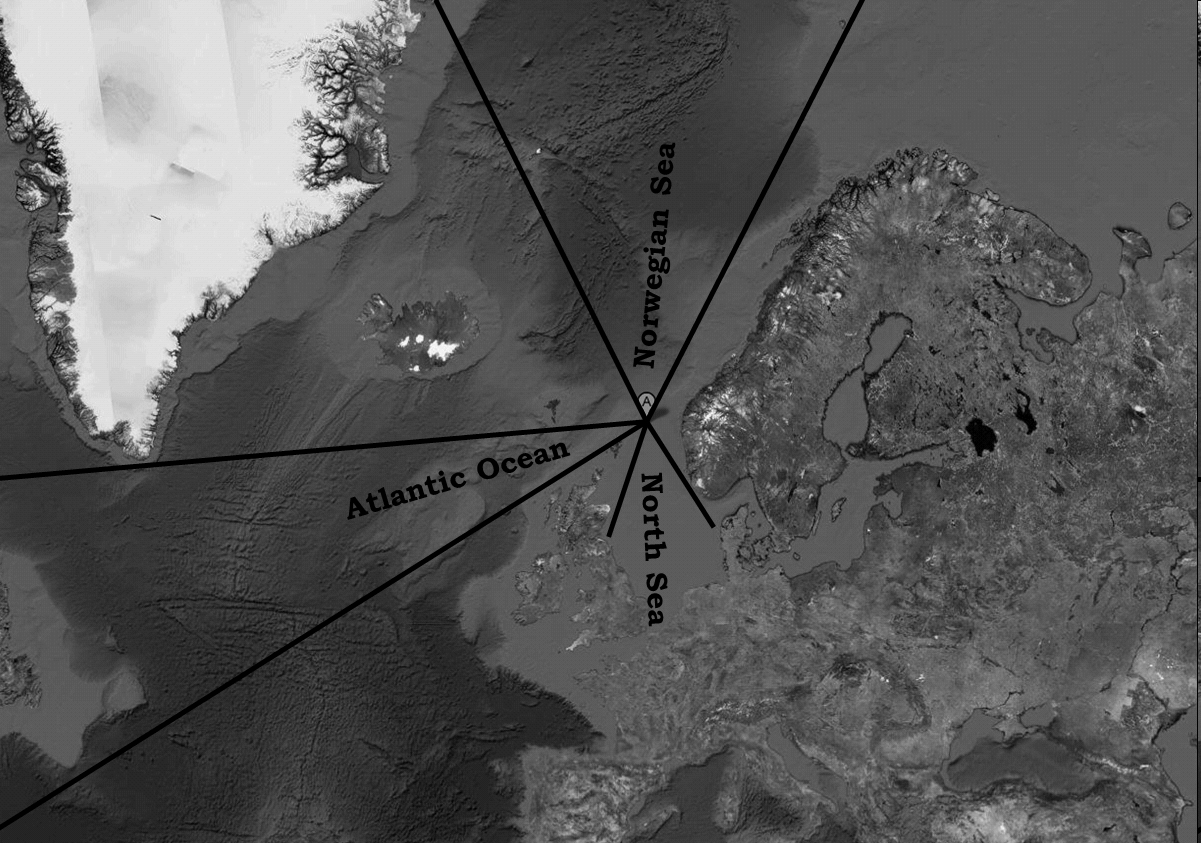}\\
  \caption{Northern North Sea location and directional sectors with distinctive wave characteristics.}
  \label{Fig4}
\end{figure}

\begin{figure}[p]
  \center\includegraphics[width=1\textwidth]{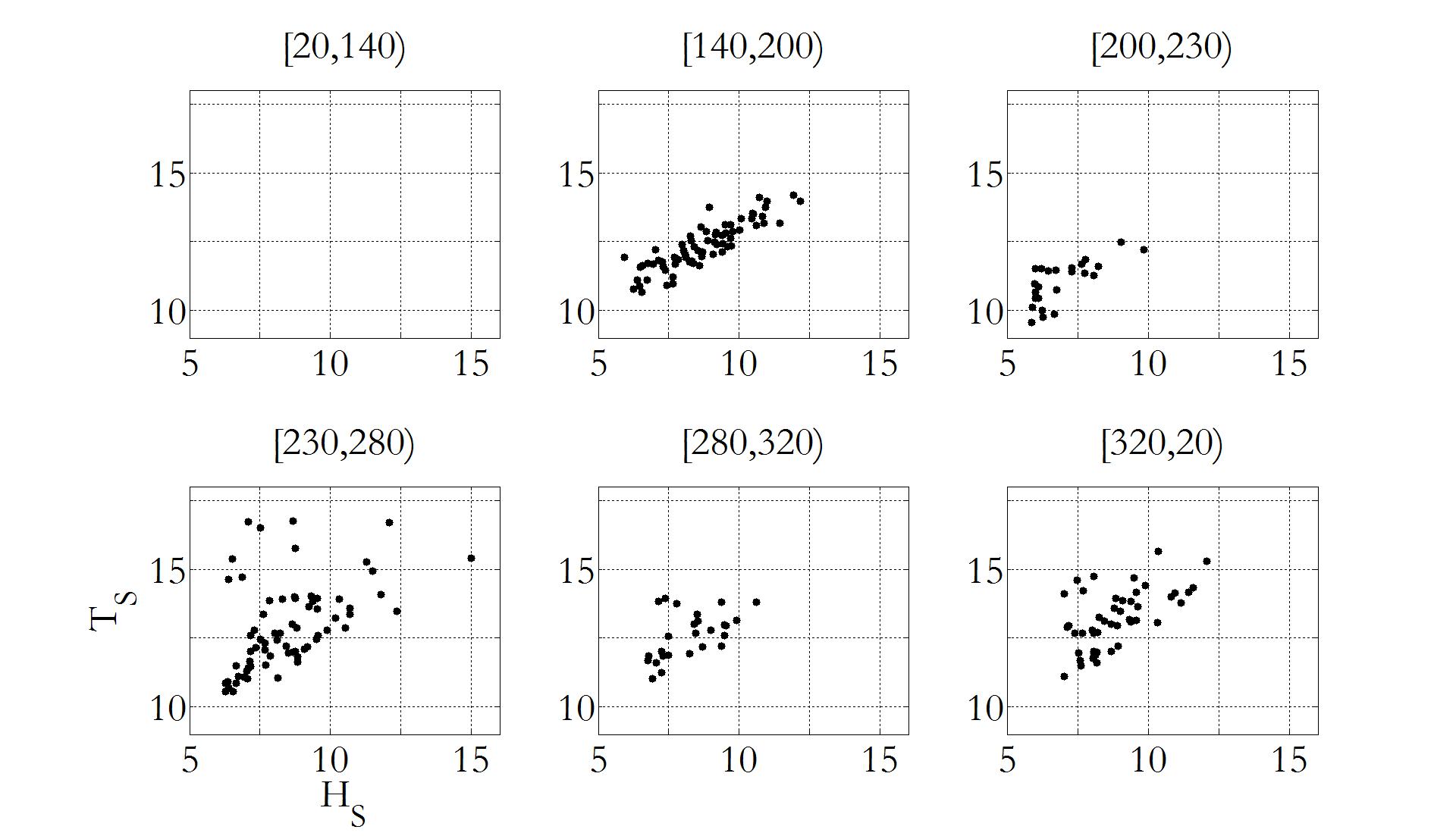}\\
  \caption{Scatter plots of associated $T_P$ against storm peak $H_S$ for different directions of arrival.}
  \label{Fig5}
\end{figure}

As an example of its application we give the results for hindcast storm peak $H_S$ and associated $T_P$ in the northern North Sea reported by \cite{JntEwnRnd13}. The objective is to model the distribution of $T_P$ for large storm peak $H_S$ as a function of storm direction. The location is particularly useful for application of the model as the wave field has identifiable characteristics for various directional sectors, as can be seen in Figure \ref{Fig4} and Figure \ref{Fig5}. Storms with the largest sea states are those occurring in the north, south, and southwest-west sectors; less severe sea states are associated with storms from the northwest sector; and virtually no storms occur that cause waves from the easterly sector. Further, it can be seen in Figure \ref{Fig5} that storm peak sea states from the northwest and southwest-west sectors are associated with the longest $T_P$ values. These characteristics should be evident in the conditional extremes modelling and can serve as a indicator of the success of the modelling. The joint distribution of $H_S$ and $T_P$ below the threshold is modelled using quantile regression.

\begin{figure}[p]
  \center\includegraphics[width=1\textwidth]{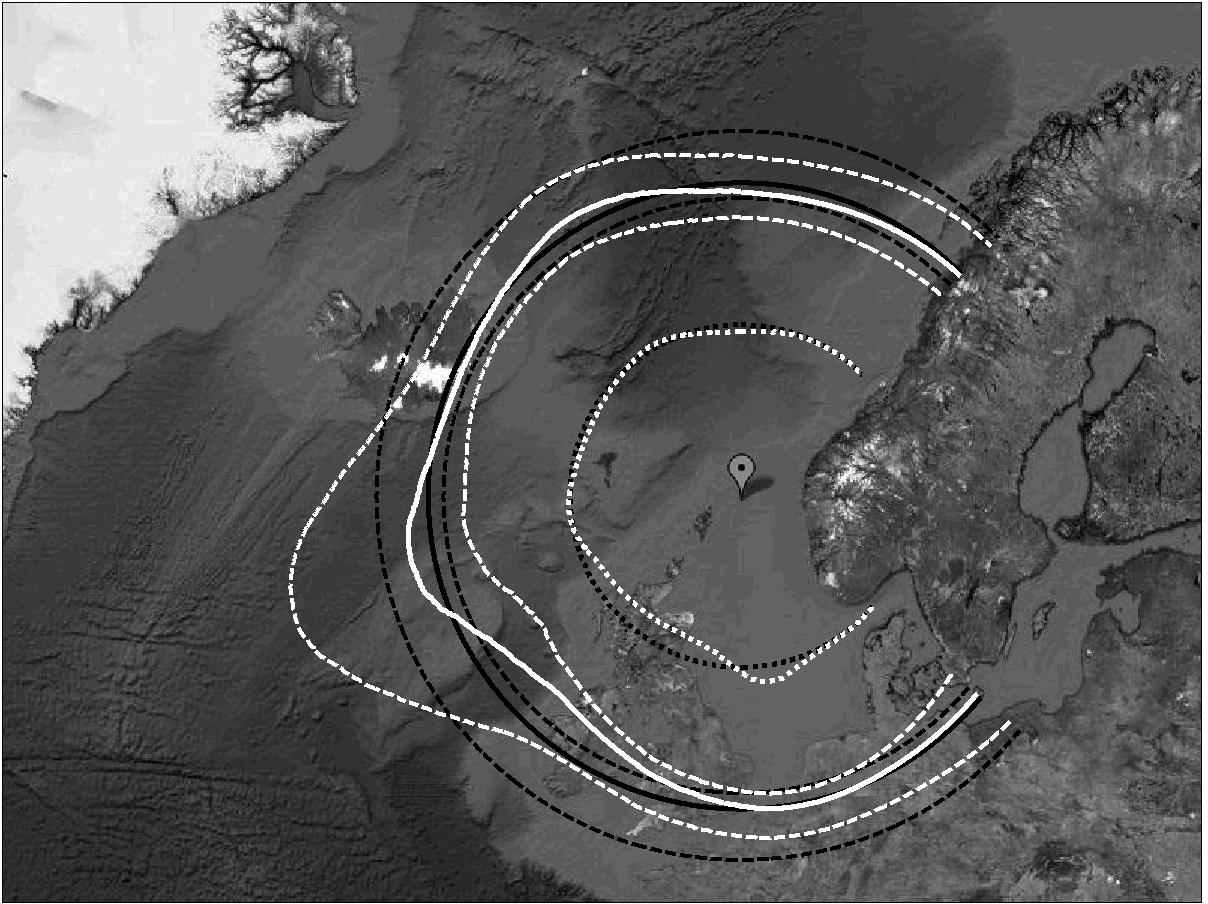}\\
  \caption{Return values of storm peak $H_S$ and associated conditional values of $T_P$. Inner dashed lines (on common scale): storm peak $H_S$ with probability of exceedence of 0.01, with (white) and without (black) directional effects. Outer solid lines (on common scale): median associated $T_P$ with (white) and without (black) directional effects; outer dashed lines give corresponding 95\% uncertainty bands for associated $T_P$.}
  \label{Fig6}
\end{figure}

Conditional $T_P$ values corresponding to storm peak $H_S$ values with exceedence probability of 0.01 are illustrated in Figure \ref{Fig6}. The inner (black and white) dotted curves, drawn on the same scale, illustrate estimates of the storm peak $H_S$ return value. The inner white dotted curve is an estimate for the directional variation of the storm peak $H_S$ return value. For comparison the inner black dotted curve is an estimate for the same return value ignoring directional effects. The influence of longer fetches from south (in particular), the Atlantic and Norwegian Sea are visible.
The outer (black and white) curves, drawn on the same scale, illustrate estimates for return values of $T_P$ conditional on exceedences of the corresponding storm peak $H_S$ value. Solid lines represent median values, and dashed lines 95\% uncertainty bands, incorporating (white) or ignoring (black) directional effects. The results clearly show increased associated periods from the Atlantic and Norwegian sectors, as expected. When directionality is ignored, associated $T_P$ values are underestimated for some sectors and overestimated for others. The importance of this difference for design can be seen in the response of a simple system with a transfer function characteristic of the roll or heave of a floating system with a natural period of around 17 seconds. Response is over-estimated by more than 30\% in directional sectors with short fetches, but under-estimated by as much as 20\% in sectors with long fetches, particularly the Atlantic sector (see \citealt{JntEwnRnd13}).

%

\section{Discussion and summary} \label{Sct4}
Estimation of joint occurrences of extremes of environmental variables is crucial for design of offshore facilities and achieving consistent levels of reliability. Specification of joint extremes in design criteria has often been somewhat ad hoc, being based on fairly arbitrary combination of extremes of variables estimated independently. Such approaches are even outlined in design guidelines. FORM and in particular inverse FORM offers the possibility for more realistic estimation of joint extremes and has been used frequently by the offshore industry since the early 90s, but FORM has generally relied on the estimation of conditional distributions developed largely from the body of the distribution rather than the tail, and it is difficult to extend to multi-dimensions. More rigourous methods for modelling joint occurrences of extremes of environmental variables are now available. In particular, the conditional extremes model and derivatives have a good theoretical basis for modelling the dependence of variables when one is extreme. These also routinely provide estimates of uncertainty, can easily be extended to multi-dimensions and to include covariates, as outlined in this paper.

Understanding extremal dependence structures is critical to reliable estimation of joint design conditions. The conditional extremes model proves a flexible framework admitting a wide range of asymptotic dependence structures described in the work of \cite{LedTwn97}, ensuring that the data (rather than unwittingly made modelling assumptions) drive the estimation of design values. There are a number of simple diagnostic tools available to characterise extremal structure, as summarised e.g. by \cite{EstKklJnt12}.

The conditional extremes model also provides a straight-forward approach to modelling of spatial extremes. In this field in particular, methods based on max-stable processes, motivated by the work of \cite{Smt90}, are increasingly popular in the statistics literature. Despite the fact that the full multivariate probability density function cannot be written in closed form, composite likelihood methods provide one approach to inference as illustrated e.g. by \cite{PdnRbtSss10}, \cite{DvsGhl11} and \cite{DvsPdnRbt12}. Censored likelihood methods provide an approach to making these models, which assume componentwise maxima, available for analysis of threshold exceedences (e.g. \cite{HsrDvs12}). Furthermore, \cite{WdsTwn12b} propose the adoption of inverted multivariate extreme value distributions with which to admit hybrid extremal dependence structures within the framework of max-stable processes. Unfortunately, approaches based on max-stable processes are methodologically complex. Much work is needed before they can be used reliably in real-world applications.
%

%
%


\pagebreak
\bibliographystyle{elsarticle-harv}
\bibliography{C:/Philip/LaTeX/Bibliography/phil}

\begin{thebibliography}{18}
\providecommand{\natexlab}[1]{#1}
\providecommand{\url}[1]{\texttt{#1}}
\expandafter\ifx\csname urlstyle\endcsname\relax
  \providecommand{\doi}[1]{doi: #1}\else
  \providecommand{\doi}{doi: \begingroup \urlstyle{rm}\Url}\fi

\bibitem[Davison and Gholamrezaee(2012)]{DvsGhl11}
A.~C. Davison and M.~M. Gholamrezaee.
\newblock Geostatistics of extremes.
\newblock \emph{Proc. R. Soc. A}, 468:\penalty0 581--608, 2012.

\bibitem[Davison et~al.(2012)Davison, Padoan, and Ribatet]{DvsPdnRbt12}
A.~C. Davison, S.~A. Padoan, and M.~Ribatet.
\newblock Statistical modelling of spatial extremes.
\newblock \emph{Statistical Science}, 27:\penalty0 161--186, 2012.

\bibitem[Eastoe et~al.(2012, under review)Eastoe, Koukoulas, and
  Jonathan]{EstKklJnt12}
E~Eastoe, S~Koukoulas, and P~Jonathan.
\newblock Statistical measures of extremal dependence illustrated using
  measured sea surface elevations from a neighbourhood of coastal locations.
\newblock \emph{Submitted to Ocean Engineering, January 2012. Draft at
  www.lancs.ac.uk/$\sim$jonathan}, 2012, under review.

\bibitem[Haver and Nyhus(1986)]{HvrNhs86}
S.~Haver and K.A. Nyhus.
\newblock A wave climate description for long term response calculations.
\newblock \emph{Proc. 5th OMAE Symp.}, IV:\penalty0 27--34, 1986.

\bibitem[Heffernan and Tawn(2004)]{HffTwn04}
J.~E. Heffernan and J.~A. Tawn.
\newblock A conditional approach for multivariate extreme values.
\newblock \emph{J. R. Statist. Soc. B}, 66:\penalty0 497, 2004.

\bibitem[Huser and Davison(2012, draft)]{HsrDvs12}
R.~Huser and A.~C. Davison.
\newblock Space-time modelling of extreme events.
\newblock \emph{Draft at arxiv.org/abs/1201.3245}, 2012, draft.

\bibitem[Jonathan and Ewans(2012, under review)]{JntEwn13}
P.~Jonathan and K.~C. Ewans.
\newblock Statistical modelling of extreme ocean environments for marine
  design.
\newblock \emph{Submitted to Ocean Engineering, July 2012. Draft at
  www.lancs.ac.uk/$\sim$jonathan}, 2012, under review.

\bibitem[Jonathan et~al.(2008)Jonathan, Ewans, and Forristall]{JntEwnFrr08a}
P.~Jonathan, K.~C. Ewans, and G.~Z. Forristall.
\newblock Statistical estimation of extreme ocean environments: The requirement
  for modelling directionality and other covariate effects.
\newblock \emph{Ocean Eng.}, 35:\penalty0 1211--1225, 2008.

\bibitem[Jonathan et~al.(2010)Jonathan, Flynn, and Ewans]{JntFlnEwn10}
P.~Jonathan, J.~Flynn, and K.~C. Ewans.
\newblock Joint modelling of wave spectral parameters for extreme sea states.
\newblock \emph{Ocean Eng.}, 37:\penalty0 1070--1080, 2010.

\bibitem[Jonathan et~al.(2012)Jonathan, Ewans, and Flynn]{JntEwnFln12}
P.~Jonathan, K.~C. Ewans, and J.~Flynn.
\newblock Joint modelling of vertical profiles of large ocean currents.
\newblock \emph{Ocean Eng.}, 42:\penalty0 195--204, 2012.

\bibitem[Jonathan et~al.(2012, draft)Jonathan, Ewans, and Randell]{JntEwnRnd13}
P.~Jonathan, K.~C. Ewans, and D.~Randell.
\newblock Joint modelling of environmental parameters for extreme sea states
  incorporating covariate effects.
\newblock \emph{In preparation for Ocean Engineering. Draft at
  www.lancs.ac.uk/$\sim$jonathan}, 2012, draft.

\bibitem[Ledford and Tawn(1997)]{LedTwn97}
A.~W. Ledford and J.~A. Tawn.
\newblock Modelling dependence within joint tail regions.
\newblock \emph{J. R. Statist. Soc. B}, 59:\penalty0 475--499, 1997.

\bibitem[Padoan et~al.(2010)Padoan, Ribatet, and Sisson]{PdnRbtSss10}
S.~A. Padoan, M.~Ribatet, and S.~A. Sisson.
\newblock Likelihood-based inference for max-stable processes.
\newblock \emph{J.~Am.~Statist.~Soc.}, 105:\penalty0 263--277, 2010.

\bibitem[Smith(1990)]{Smt90}
R.~L. Smith.
\newblock Max-stable processes and spatial extremes.
\newblock \emph{Unpublished, available from
  http://www.stat.unc.edu/postscript/rs/spatex.pdf}, 1990.

\bibitem[Tromans and Vanderschuren(1995)]{TrmVnd95}
P.~S. Tromans and L.~Vanderschuren.
\newblock Risk based design conditions in the {N}orth {S}ea: Application of a
  new method.
\newblock \emph{Offshore Technology Confernence, Houston (OTC--7683)}, 1995.

\bibitem[Wadsworth and Tawn(2012)]{WdsTwn12b}
J.L. Wadsworth and J.A. Tawn.
\newblock Dependence modelling for spatial extremes.
\newblock \emph{Biometrika}, 99:\penalty0 253--272, 2012.

\bibitem[Winterstein and Engebretsen(1998)]{WntEng98}
S.~Winterstein and K.~Engebretsen.
\newblock Reliability-based prediction of design loads and responses for
  floating ocean structures.
\newblock In \emph{Proc. 27th International Conf. on Offshore Mechanics and
  Arctic Engineering, Lisbon, Portugal}, 1998.

\bibitem[Winterstein et~al.(1993)Winterstein, Ude, Cornell, Bjerager, and
  Haver]{WntEA93}
S.~R. Winterstein, T.~C. Ude, C.~A. Cornell, P.~Bjerager, and S.~Haver.
\newblock Environmental parameters for extreme response: Inverse \textsc{Form}
  with omission factors.
\newblock In \emph{Proc. 6th Int. Conf. on Structural Safety and Reliability,
  Innsbruck, Austria}, 1993.

\end{thebibliography}
\end{document}